\documentclass{article}

\usepackage{PRIMEarxiv}

\usepackage[utf8]{inputenc} 
\usepackage[T1]{fontenc}    
\usepackage{hyperref}       
\usepackage{url}            
\usepackage{booktabs}       
\usepackage{nicefrac}       
\usepackage{microtype}      
\usepackage{lipsum}
\usepackage{fancyhdr}       
\usepackage{graphicx}       
\graphicspath{{media/}}     
\usepackage{subcaption}

\usepackage{amsmath,amssymb,amsfonts}
\usepackage{algorithmic}
\usepackage{textcomp}
\usepackage{xcolor}
\usepackage[numbers,square]{natbib}
\usepackage{stfloats,framed}
\usepackage{geometry}
\usepackage{authblk}

\title{Automatic Contextual Audio Denoising}

\author[1,2]{ Diep Luong}
\author[2]{Konstantinos Drossos}
\author[2]{Mikko Heikkinen}
\author[1]{Tuomas Virtanen}

\affil[1]{Tampere University, Finland}
\affil[2]{Nokia, Finland}

\begin{document}

\twocolumn[
\maketitle 
\begin{abstract}
Audio context determines which sound components and sources are relevant and which can be perceived as irrelevant (noise) by listeners. For example, traffic noise is informative in urban surveillance but noise for a phone call at the same location. Most current audio denoising systems apply fixed target–noise definitions, often removing useful components in one context while failing to suppress irrelevant components. To address this, we introduce the concept \emph{automatic contextual audio denoising} (ACAD) which defines target and noise based on the inferred context. In this work, we restrict context to be associated with an acoustic scene class. We label sound events outside the event distribution of a scene class (noise) as out-of-context (OC) and events typical for that scene as in-context (IC). We implement a deep learning method that automatically infers the context of the audio signal and removes OC components, and benchmark it against variants: without context inference, with oracle context, and with separately provided uninformative context. On paired clean/noisy data across diverse contexts, where OC components in one context may be IC in another, our proposed method outperforms other approaches across standard objective metrics, indicating that the model can infer context and context-dependent processing can enhance denoising.
\vspace{0.15in}
\end{abstract}
]

\keywords{automatic contextual audio denoising, audio denoising, source separation, context-dependent, audio understanding, neural network.}

\section{Introduction}
The distinction between relevant and irrelevant (noise) sound components and sources inherently depends on different audio contexts and the listener’s goals~\cite{gaver:1993:ecological}. For example, traffic noise is an informative source in urban surveillance or traffic monitoring systems but noise for a phone call on the street at the same location. Since denoising aims at suppressing components that interfere with the source of interest, the objective of audio denoising systems should extend beyond merely estimating a fixed, predefined target source from a noisy audio mixture. To optimize the overall perceptual listening experience, it is beneficial to define target and noise components depending on the audio context.

Despite impressive advancements of deep learning approaches in audio denoising~\cite{drossos:2018:ijcnn,aubreville:2018:iwaenc,giri:2019:waspaa,moliner:2022:icassp,drossos:2025:mmsp}, the vast majority of current models falls short of considering the audio context. Typically, models are trained to learn a mapping from noisy to target signal, where the targets are specific classes of sound (e.g. speech). Under reconstruction‑based objectives, models learn global, fixed inference rules that distinguish target and noise components based on the statistical characteristics learned during training. A rigid definition of target components and the lack of context dependence most likely strips away from the audio components and sources that are useful in some context, which contribute to an immersive and natural audio experience, while fails to suppress irrelevant components.

Related work on source separation offers mechanisms that enable more adaptive definition of target sources. Instead of relying on a single global definition of the target across all mixtures, target source separation methods condition the model on information about the target source to determine the target distribution. Enrollment audio is often used for conditioning in target speaker extraction~\cite{zmolikova:2019:jstsp, zhang:2025:icassp}. Beyond audio guidance, several studies leverage visual information about the target speaker to improve separation~\cite{ochiai:2019:interspeech, afouras:2019:interspeech, li:2020:icassp}. 
Multiple cross-modality cues (e.g., sound event tags, textual descriptions, and video clips) have been employed for target source extraction~\cite{li:2023:icassp}.
Text-derived semantic information of the target sources has also been used in extraction task~\cite{chen25l:2025:interspeech}. Besides multimodal information, human-input guidance via text-based input prompts has been utilized to condition the separation~\cite{hao:2025:tcds}. In all these studies, auxiliary information of the target source is required to guide the separation process.

In other audio processing tasks, context information has been used to enhance the performance. For example, predefined scene contexts have been used to facilitate sound event detection in seen and unseen scenes~\cite{tonami:2022:icassp}. Taking into account that what a person says depends on the current particular context they are in, a contextualized automatic speech recognition model~\cite{pundak:2018:slt} has been proposed.

To address the lack of context dependence in current audio denoising and motivated by the use of context in other audio processing tasks, we propose the concept of \textbf{automatic contextual audio denoising} (ACAD). In ACAD, target and noise are context-dependent, i.e. ACAD defines the target and noise distributions under the learned context, and different contexts can give different definitions of target and noise. ACAD hence avoids static, context-independent definitions of target and noise, and operates without explicit guidance about which sound components constitute the target.

Under the above general definition of ACAD, in this study, we focus on removing out-of-context (OC) components from a noisy mixture based on the learned audio context. We define OC components that can detract from the primary audio context as noise, while such OC components in one context can be in-context (IC) in another. To start and foster further work on ACAD, we present a first baseline method and a corresponding dataset~\cite{luong:2026:zenodo} consisting of pairs of clean and noisy audio across different contexts, which we release publicly\footnote{\url{https://doi.org/10.5281/zenodo.20287453}}. The baseline is a DNN-based model that infers the context and performs denoising using only the noisy audio signal as input. We evaluate the performance of the baseline method against three other approaches, with no context inference, with oracle context, and with separately provided uninformative context. Obtained results show that the proposed ACAD method outperforms the other approaches, suggesting that informative context can enhance the ACAD process and that the learned context (compared to oracle-based approach) can further enhance the ACAD. The rest of the paper is organized as follows: Section~\ref{sec:method} presents the problem setup, the dataset, and our proposed method. In Section~\ref{sec:evaluation} we present the evaluation setup of the ACAD method. Results are discussed in Section~\ref{sec:results}, and Section~\ref{sec:conslusion} concludes this paper. 
\section{Problem setup, dataset, \& proposed method}
\label{sec:method}
In ACAD, a method takes a single input audio signal, $\tilde{\mathbf{x}}$, infers its context, $\mathbf{e}$, and outputs $\mathbf{x}$ in which all components and sources considered noise with respect to $\mathbf{e}$ are removed.

In this first study, we consider each audio context to be associated with one acoustics scene class, and we use a fixed set of acoustic scene classes. We define context as the statistical properties of a audio signal, that can consistently be classified with a specific acoustic scene class. Such statistical properties can include, but are not limited to, sound event distribution, spectral and temporal patterns, loudness, dynamic range, and other related and high-level factors that can be used to define a scene. OC component (or noise) denotes outlier sound events that do not belong to the current scene’s sound event distribution. 

Consequently, we consider $\tilde{\mathbf{x}}$ as the noisy audio signal and $\mathbf{x}$ as the corresponding clean one without any OC components under the current context. Thus, the ACAD method takes as an input $\tilde{\mathbf{x}}$, infers $\mathbf{e}$, and performs contextual denoising on $\tilde{\mathbf{x}}$. The output of the method is an estimate of $\mathbf{x}$, $\hat{\mathbf{x}}$.  
\subsection{Dataset construction}
\label{sec:method:data}
To create our ACAD dataset, we employ a dataset of real-life recordings from different acoustic scene classes and a labeled sound event dataset. The former serves as the clean (target) signals $\mathbf{x}$ and the latter as the pool for the OC components used to construct noisy scenes $\tilde{\mathbf{x}}$ by mixing them with the clean signals. The acoustic scene audio from CochlScene dataset~\cite{jeong:2022:apsipa} and the OC components from FSD50K~\cite{fonseca:2021:ieee} are used to create the dataset in this study. We focus on six acoustic scene classes from CochlScene: Kitchen, Park, Restaurant, Restroom, Street, and Subway, which are distinct from each other considering their audio content. We construct the noisy audio by mixing clean audio and OC events as noise.

\begin{figure}
    \centering
    \includegraphics[width=0.95\columnwidth]{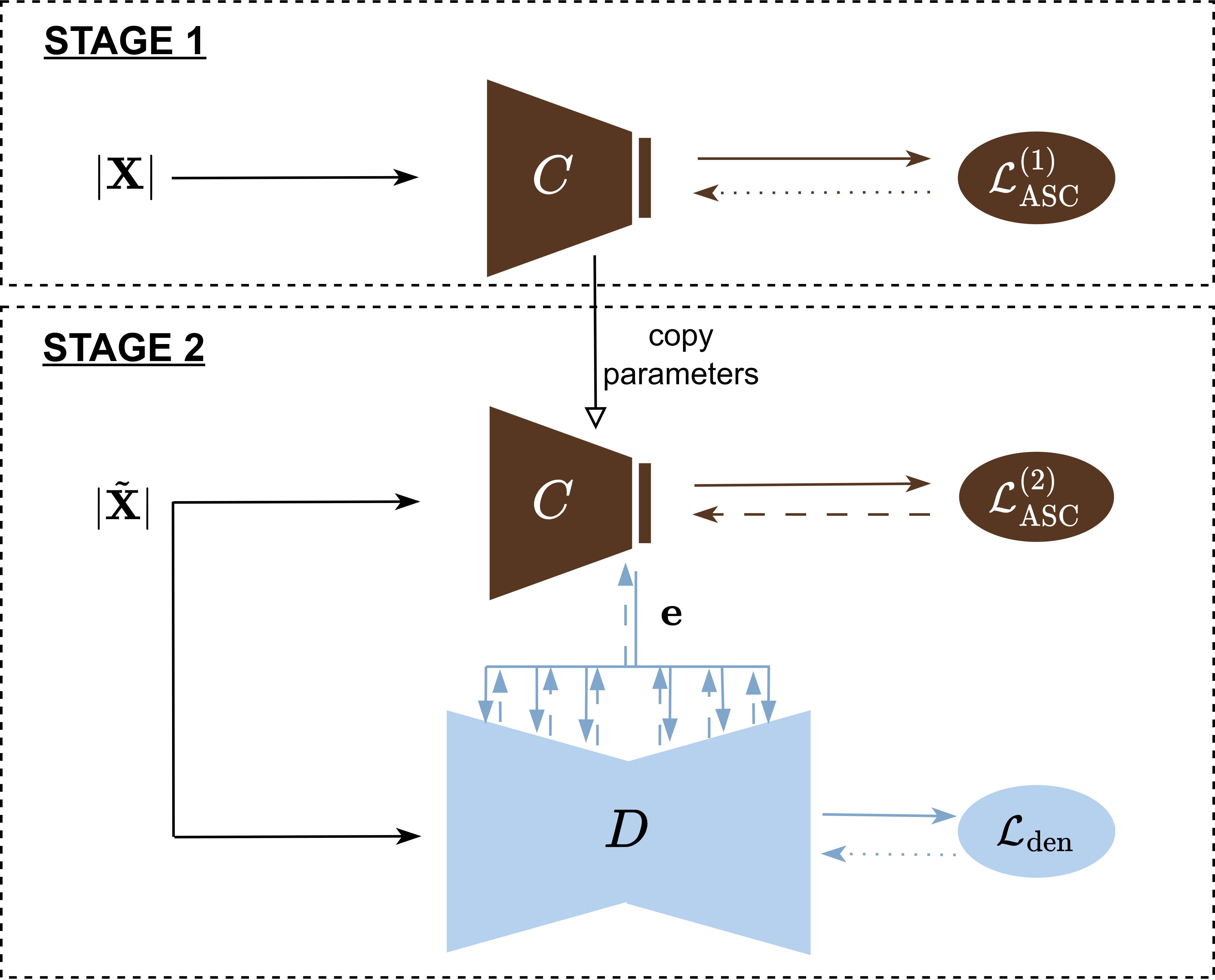}
    \caption{Overview of our ACAD method. Context extractor $C$ is pretrained for ASC on clean audio spectrogram. Denoiser $D$ removes OC components from noisy spectrogram under learned context $\mathbf{e}$ from $C$. In the frozen setting, $C$ is fixed; in the finetuned setting, $C$ is trained jointly with $D$. Backpropagation is shown with dashed or dotted arrows; backpropagation along dashed arrows is only available in the finetuned setting.}
    \label{fig:contextual-denoising}
\end{figure}

To identify OC components for each acoustic scene class, we first define preliminary IC and OC sets of event classes. A set of event classes within each scene class is identified using pre-trained PANNs for sound event detection~\cite{kong:2020:ieee}. Within each time step in the PANNs output, if an event label and any of its child labels in the AudioSet ontology~\cite{gemmeke:2017:icassp} are present together, the parent label is removed while retaining all child labels (e.g., if “Animal”, “Bird”, and “Dog” are present, only “Bird” and “Dog” are kept). Event classes within one scene class are sorted in a decreasing order according to their activity, which is quantified as the total active duration. The 20 most active event classes and their hierarchical descendants are considered as the  preliminary IC set for each scene class. Conversely, the direct siblings (and their descendants) of the 20 most active event classes form the preliminary OC. We augment the preliminary OC set with the 20 most active events (with descendants) and their siblings (with descendants) from other scene classes, but excluding those are already in the preliminary IC set.

A manual refinement step is  performed to ensure a clear distinction between IC and OC events. This involved informal human judgement on the semantic similarity between each preliminary IC event class, $k$, and any of its ancestors, $p$, found in the the preliminary set of OC event classes, following three rules: i) \textit{Highly similar parent}: if $p$ and its descendants are very similar to $k$, $p$ and descendants are discarded from OC, 2) \textit{Partially similar parent}: if $p$ has diverse descendants distinct from $k$ but $p$ itself could sound similar to $k$, only $p$ is removed from OC, and 3) \textit{Distinct parent}: if $p$ and its descendants are diverse and acoustically distinct from $k$, $p$ remains in OC. 

The dataset construction process ensures that the OC to be removed in a specific scene class can appear to be the IC within other acoustic scene classes. This is an important point for the ACAD task, since it enforces context-dependent learning of OC components and impede the model to rigidly learn components to be removed globally across all scene classes. 

Specifically, following the steps mentioned above, we employ Scaper library~\cite{salamon:2017:waspaa} to generate $N_{train}=10000$ noisy-clean pairs of 10 sec audio for training and $N_{val/test}=3000$ for validation/testing for each of the ($C=6$) acoustic scene classes. We follow the splitting of the employed datasets, mixing scenes and events from corresponding splits (e.g., training with training, etc). Each clean audio file, scaled to a random LUFS level in [-10,-15] LUFS, is mixed with 1 to 3 OC event classes selected from the refined OC set. OC events, with 1 to 2 instances per event class (each of 0.5s-3s long), are added to the clean audio at a level of [-5,10] dB compared to the background scene.

Finally, we obtain
\begin{equation*}
  \mathbb{D}=\{\{(\tilde{\mathbf{x}}^{n,i}, \mathbf{x}^{n,i})\}_{n=1}^N, c^{i}\}_{i=1}^{C},
\end{equation*}
\noindent
comprising of $N$ pairs of $\tilde{\mathbf{x}}$ and corresponding $\mathbf{x}$ for each of $C$ distinct acoustic scene classes $c^i$.
\subsection{Proposed method}
Our method for ACAD, illustrated in Figure~\ref{fig:contextual-denoising}, consists of a context extractor $C$ and denoising model $D$. $C$  takes audio $\tilde{\mathbf{x}}$ as an input and outputs a latent embedding $\mathbf{e}$ that represents the estimated denoising context. $D$ takes $\tilde{\mathbf{x}}$ as an input, conditioned on  $\mathbf{e}$, and outputs $\hat{\mathbf{x}}$. We employ a two-stage training for our method, including first the pretraining of $C$ followed by the training of $D$. In the second stage, we integrate $C$ in two settings; i) frozen, where $C$ is kept fixed during the training of $D$, and ii) finetuned, where $C$ is finetuned jointly with $D$. 

In the first stage and in both settings, $C$ is pretrained to perform acoustic scene classification (ASC) on clean audio $\mathbf{x}$. Specifically, $C$ computes the log Mel spectrogram $|\mathbf{X}|_\text{mel}$ by applying Mel filter banks to the short-time Fourier transform (STFT) magnitude spectrogram $|\mathbf{X}|$ of $\mathbf{x}$, and outputs the probability distribution of acoustic scene classes. For the pretraining of $C$, we minimize the cross-entropy loss for ASC
\begin{equation}
    \mathcal{L}_\text{ASC} = \mathbb{E}_{(\mathbf{x}, c)\sim\mathbb{D}} \left[-\log{p_c}\right]
\end{equation}
where $p_c$ is the predicted probability corresponding to the ground truth acoustic scene class $c$.

In the second stage, $D$ is trained to do denoising of $\tilde{\mathbf{x}}$, given the inferred context. To condition $D$ on the context, $C$ computes $|\tilde{\mathbf{X}}|_\text{mel}$ and outputs a latent representation, $\mathbf{e}$, which is the inferred context from $\tilde{\mathbf{x}}$. Employing similar STFT parameters to what is used for $C$, we yield the magnitude spectrum $|\tilde{\mathbf{X}}|$ and corresponding phase $\angle\tilde{\mathbf{X}}$ from $\tilde{\mathbf{x}}$. Then, $D$ takes $|\tilde{\mathbf{X}}|$ and $\mathbf{e}$ as inputs and outputs a contextual denoising mask with values in $[0,1]$. The estimated clean audio spectrogram $|\hat{\mathbf{X}}|$ is obtained by masking as
\begin{equation}
    |\hat{\mathbf{X}}| = |\tilde{\mathbf{X}}| \odot D(|\tilde{\mathbf{X}}|,\mathbf{e}).
\end{equation}
\noindent
where $\odot$ is the element-wise product. We optimize $D$ by minimizing the loss 
\begin{equation}
    \mathcal{L_\text{den}} = \mathbb{E}_{(\mathbf{x}, {\tilde{\mathbf{x}}})\sim\mathbb{D}} \left[l(\mathbf{x}, \hat{\mathbf{x}})\right]
\end{equation}
where $\hat{\mathbf{x}}$ is the estimated clean audio reconstructed from $|\hat{\mathbf{X}}|$ and $\angle\tilde{\mathbf{X}}$, $l(\cdot)$ is a reconstruction loss (e.g., $L_{p}$-based loss or typical source separation loss like [SI-]SDR/SNR). 

In the second setting, where $C$ is finetuned during the training of $D$, the optimization minimizes the joint loss
\begin{equation}
    \mathcal{L}_{\text{tot}} = \lambda_{\text{ASC}}\mathcal{L}_{\text{ASC}} + \lambda_{\text{den}}\mathcal{L}_{\text{den}},
\end{equation}
\noindent
where $\lambda_{\text{ASC}}$ and $\lambda_{\text{den}}$ are weights for each corresponding loss. 
\section{Evaluation}
\label{sec:evaluation}
\subsection{Model architectures}
In our model, $C$ is a CRNN-based architecture consisting of three convolutional blocks with residual connections, followed by an recurrent neural network (RNN) and temporal attention pooling. Each convolutional block contains a convolutional neural network (CNN) with a square kernel of $3\times3$ with $2\times2$ stride, doubling the channel dimension, and a $3\times3$ CNN with unit stride. The later CNN preserves the feature's dimension, and a residual connection is employed between its input and output. The first block outputs 8 channels. The final CNN's output is processed by an RNN with 128 hidden units, followed by temporal attention pooling and two fully connected layers of sizes 64 and the number of scene classes. An intermediate embedding is extracted after the first fully connected layer of $C$, serving as context $\mathbf{e}$.

Our $D$ is a UNet with skip-connections, following a 3-block deep encoder-decoder structure. Each encoder block consists of a CNN with a $3\times3$ kernel, doubling the number of channels at each successive layer. The initial encoder layer outputs 16 channels. The bottleneck layer employs a CNN with a $3\times3$ kernel and a transposed CNN (TrCNN) layer with the same kernel size. In the decoder path, following the concatenation of features from the skip connections, each decoder block utilizes a CNN with a $3\times3$ kernel and unit stride to halve the channel dimension and a TrCNN for upsampling. All CNNs and TrCNNs, apart the CNNs in the decoder, use a stride of $2\times2$. The final output of the UNet is a mask, applied to $|\tilde{\mathbf{X}}|$ to obtain $|\hat{\mathbf{X}}|$. To incorporate the context during the training of $D$, we leverage feature-wise linear modulation (FiLM) layers~\cite{perez:2018:aaai} with $\mathbf{e}$ at every layer within the encoder and decoder of $D$ to allow modulating the feature maps at multiple scales.
\subsection{Training setup}
Audio signals are resampled to 22050 kHz to balance computational cost and retained signal information. We employ a STFT with window size of 1024 samples, 50\% overlap, and a 64-band Mel filter bank over the 10-sec audio to extract the input features into model. For training, we used a batch size of 64 and Adam optimizer, with default hyper-parameters and learning rate of $10^{-3}$. SI-SNR loss is used to for the reconstruction loss. We use $\lambda_{ASC}=\lambda_{den}=1$.
\subsection{Baselines}
We use a common UNet backbone for the denoising model $D$ and perform an ablation of context utilization strategies. Specifically, we compare our our method, $\text{UNet}_{\text{ASC}}$, against three variants: i) without any context inference indicated as UNet, ii) with oracle context indicated as $\text{UNet}_{\text{oracle}}$, and iii) with separately provided uninformative context indicated as $\text{UNet}_{\text{const}}$. UNet model is a standard UNet architecture without FiLM layers and with no context inference from the context extractor $C$; this serves as a naive baseline. To investigate the effect of conditioning on the oracle context (scene class), we employ $\text{UNet}_{\text{oracle}}$ and feed the FiLM layers with the oracle one-hot encoded scene class. At $\text{UNet}_{\text{const}}$, uninformative context is provided through a constant vector of all ones as the conditioning vector. For $\text{UNet}_{\text{ASC}}$, there are two variants, $\text{UNet}_{\text{Fr-ASC}}$ and $\text{UNet}_{\text{Tu-ASC}}$,
depending on whether $C$ is frozen or finetuned during denoising. To ensure a robust comparison, we experiment with the $\text{UNet}_{\text{const}}$ using different embedding sizes for the conditioning vector, matching those used for $\text{UNet}_{\text{ASC}}$ (\emph{Embedding size I}) and $\text{UNet}_{\text{oracle}}$ (\emph{Embedding size II}). Different experiment setups along with their conditioning inputs are summarized in Table~\ref{tab:models}.
\begin{table}[t]
    \centering
    \normalsize
    \caption{Summary of different experiment setups}
    \label{tab:models}
    \begin{tabular}{ll}
    \textbf{Model} & \textbf{Context utilization} \\
    \hline
    UNet & None\\
    $\text{UNet}_{\text{ASC}}$ & ASC embedding \\
    $\text{UNet}_{\text{oracle}}$ & Oracle scene class \\
    $\text{UNet}_{\text{const}}$ & Uninformative constant vector \\
    \end{tabular}
\end{table}
\begin{table}[t!]
    \centering
    \normalsize
    \caption{Mean/STD of evaluation metrics for the proposed ACAD method. \emph{Embedding size} refers to the conditioning vector's shape. \emph{Embedding size I} is (batch, ASC emb size) and \emph{Embedding size II} is (batch, num scene class)}
    \label{tab:results}
    \begin{tabular}{lcc}
    \textbf{Model} & \textbf{SI-SDR (dB)} & \textbf{SDR (dB)}\\
    \hline
    Noisy input & 4.27/0.00 & 4.26/0.00\\
    UNet & 10.16/0.02 & 10.56/0.02 \\
    \hline
    \textit{Embedding size I} & &\\
    \quad $\text{UNet}_{\text{Tu-ASC}}$ & \textbf{12.12/0.04} & \textbf{12.56/0.04}\\
    \quad $\text{UNet}_{\text{Fr-ASC}}$ & 11.04/0.07 & 11.47/0.09\\
    \quad $\text{UNet}_{\text{const}}$ & 10.02/0.01 & 10.41/0.03\\
    \hline
    \textit{Embedding size II} & &\\
    \quad $\text{UNet}_{\text{oracle}}$ & 10.82/0.02& 11.23/0.03\\
    \quad $\text{UNet}_{\text{const}}$ & 10.13/0.03& 10.53/0.05\\
    \end{tabular}
\end{table}
\section{Results and discussion}\label{sec:results}
Table~\ref{tab:results} presents the results of all experiments. We report the mean and standard deviation (STD) of the scale-invariant signal-to-distortion ratio (SI-SDR) and signal-to-distortion ratio (SDR) between the estimated signal and the clean ground truth obtained through five runs with identical experimental settings for each experiment. The result is calculated on the testing split of our dataset.

As shown in Table~\ref{tab:results}, UNet, serving as the lower bound reference for evaluation, yields 10.16 dB SI-SDR and 10.56 dB SDR. With oracle context, $\text{UNet}_{\text{oracle}}$ achieves a 0.66 dB and 0.67 dB gain in SI-SDR and SDR over UNet. Learned context offers even larger improvements. A possible explanation is that the oracle context is constrained to scene-class information, whereas the vector-based representation learned from the noisy audio provides richer contextual information. $\text{UNet}_\text{Fr-ASC}$ and $\text{UNet}_\text{Tu-ASC}$ employs the same context extractor $C$ from the first training stage, which obtains an accuracy of 84.18\% in ASC on the testing set. Using the learned context, $\text{UNet}_\text{Fr-ASC}$ observes increases of 0.88 dB in SI-SDR and 0.91 dB in SDR over UNet. Further finetuning $C$ jointly with the denoising model $D$ in $\text{UNet}_\text{Tu-ASC}$ yields additional gains, outperforming UNet by 1.96 dB and 2 dB in SI-SDR and SDR, respectively. This suggests finetuning allows better alignment between the context extractor $C$ and the denoiser $D$ for denoising. In contrast, both $\text{UNet}_{\text{const}}$ variants, with uninformative context, show inferior performance compared UNet. This indicates that uninformative information does not help or might even distract the denoising task. Although the unconditioned UNet in principle can internally infer the context during denoising, incorporating informative context, either via learned context ($\text{UNet}_{\text{ASC}}$) or through oracle context ($\text{UNet}_{\text{oracle}}$), shows consistent improvement in the denoising performance compared to the model without context inference.

\begin{figure}[t!]
    \centering
    \begin{subfigure}[b]{0.4\columnwidth}
        \centering
        \includegraphics[width=\textwidth]{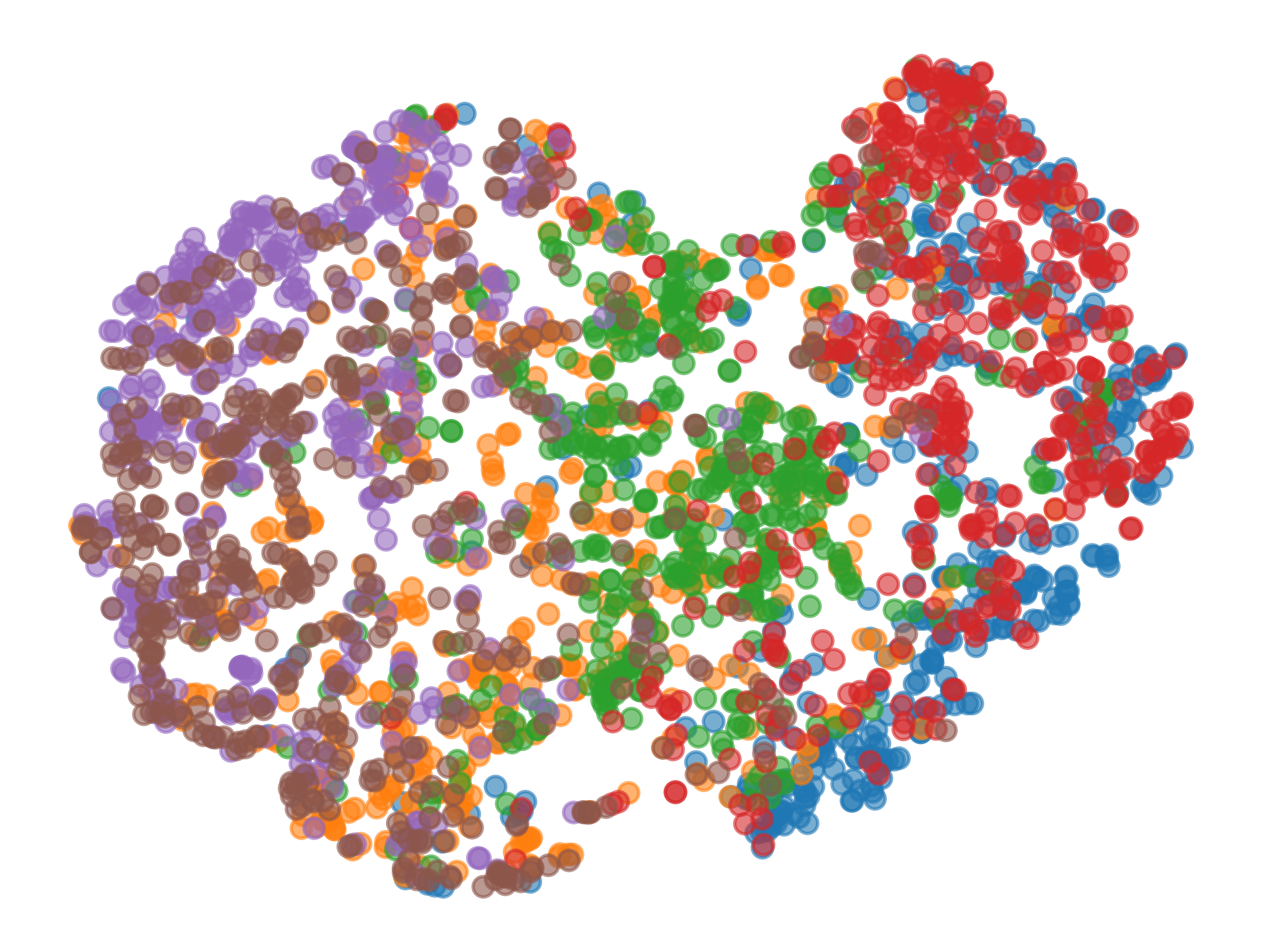}
        \caption{}
    \end{subfigure}
    \begin{subfigure}[b]{0.4\columnwidth}
        \centering
        \includegraphics[width=\textwidth]{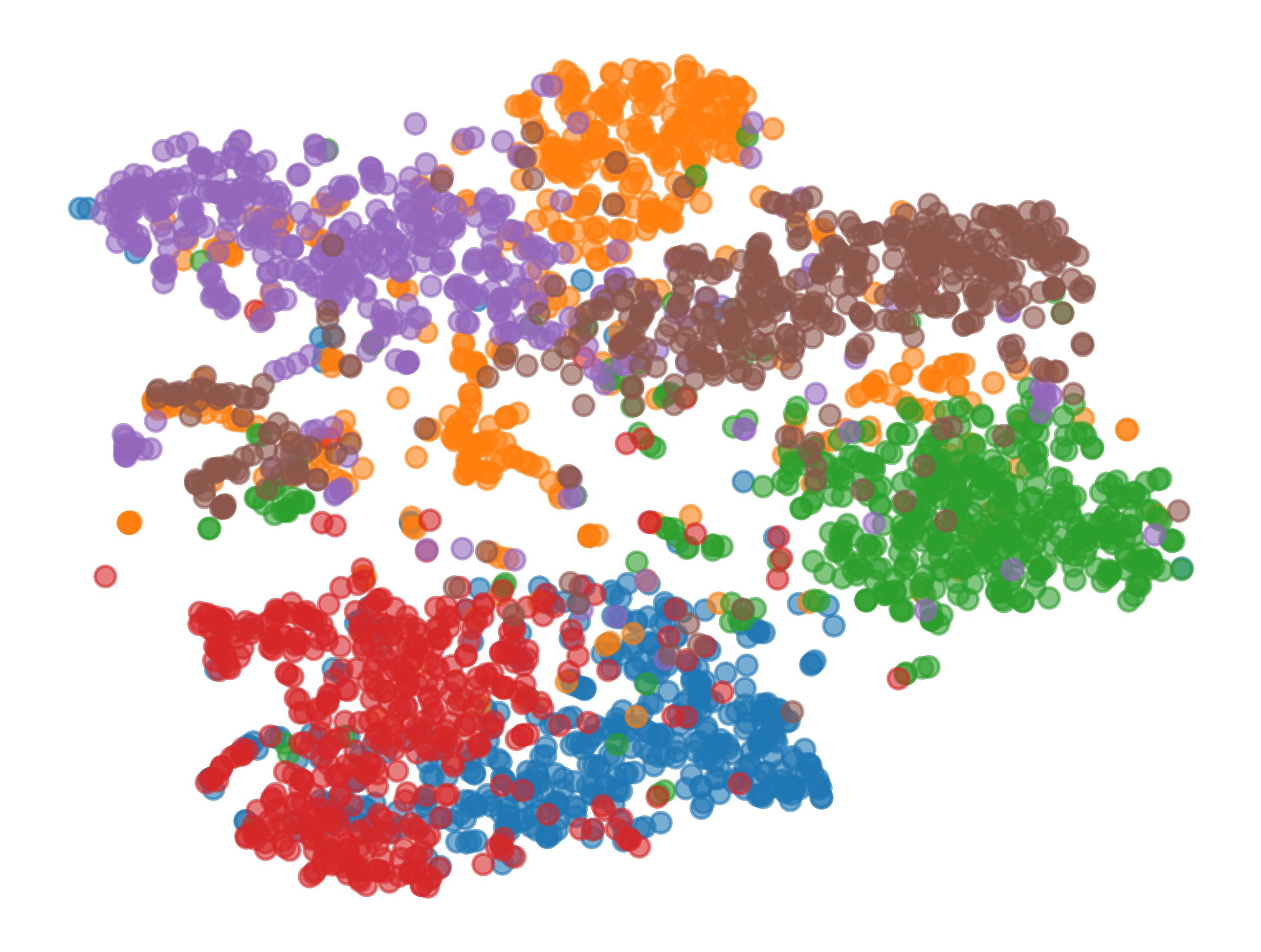}
        \caption{}
    \end{subfigure}
    \begin{subfigure}[b]{0.4\columnwidth}
        \centering
        \includegraphics[width=\textwidth]{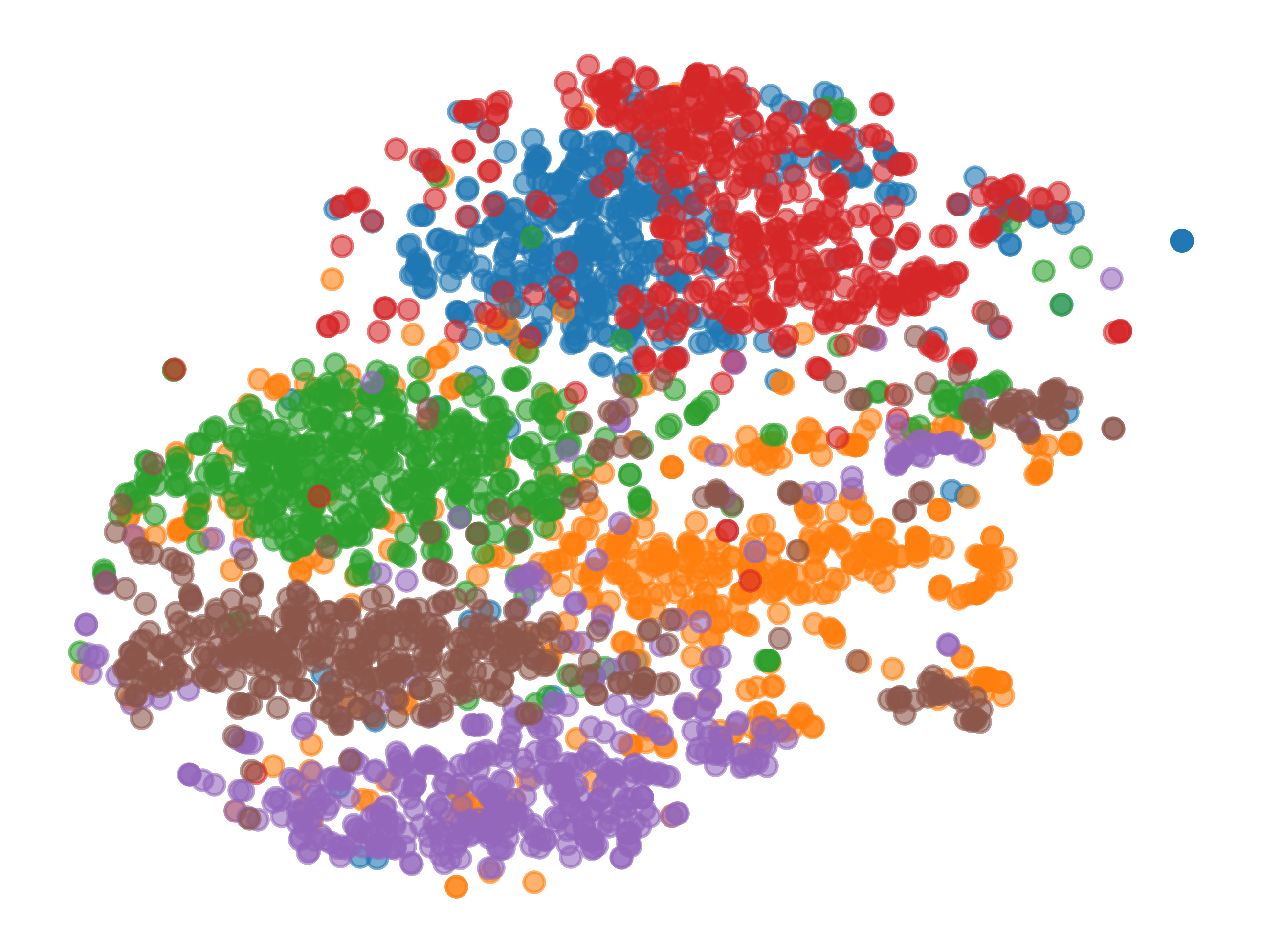}
        \caption{}
    \end{subfigure}
    \begin{subfigure}[b]{0.4\columnwidth}
        \centering
        \includegraphics[width=1.2\textwidth]{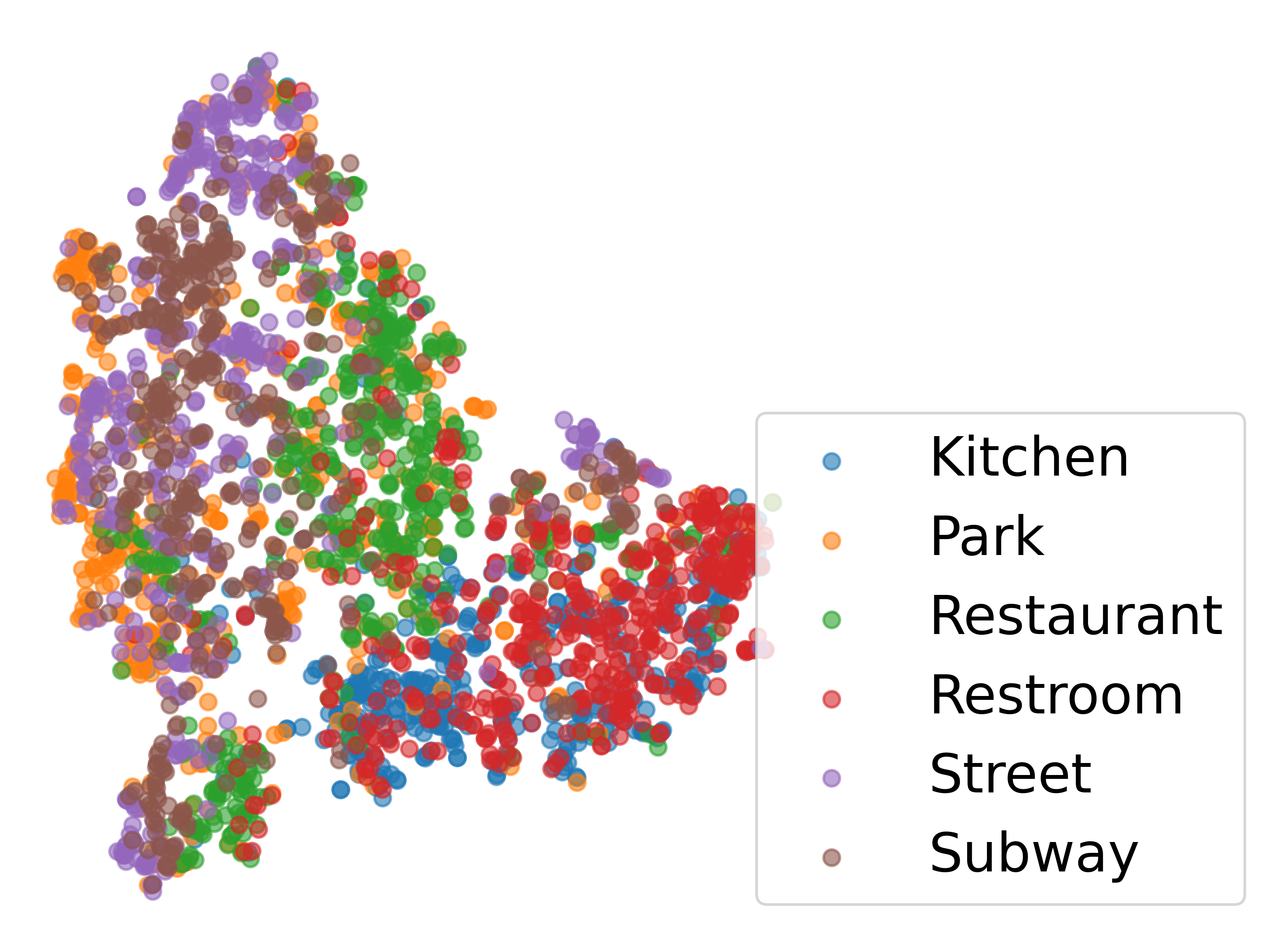} 
        \caption{}
    \end{subfigure}
    \caption{t-SNE representation of the bottleneck features from $D$ of (a) UNet, (b) $\text{UNet}_{\text{Fr-ASC}}$, (c) $\text{UNet}_{\text{Tu-ASC}}$, and (d) $\text{UNet}_{\text{const}}$ (\emph{Embedding size I})}
    \label{fig:filmunet-asc}
\end{figure}
\begin{figure}[t!]
    \centering
    \begin{subfigure}[b]{0.4\columnwidth}
        \centering
        \includegraphics[width=\textwidth]{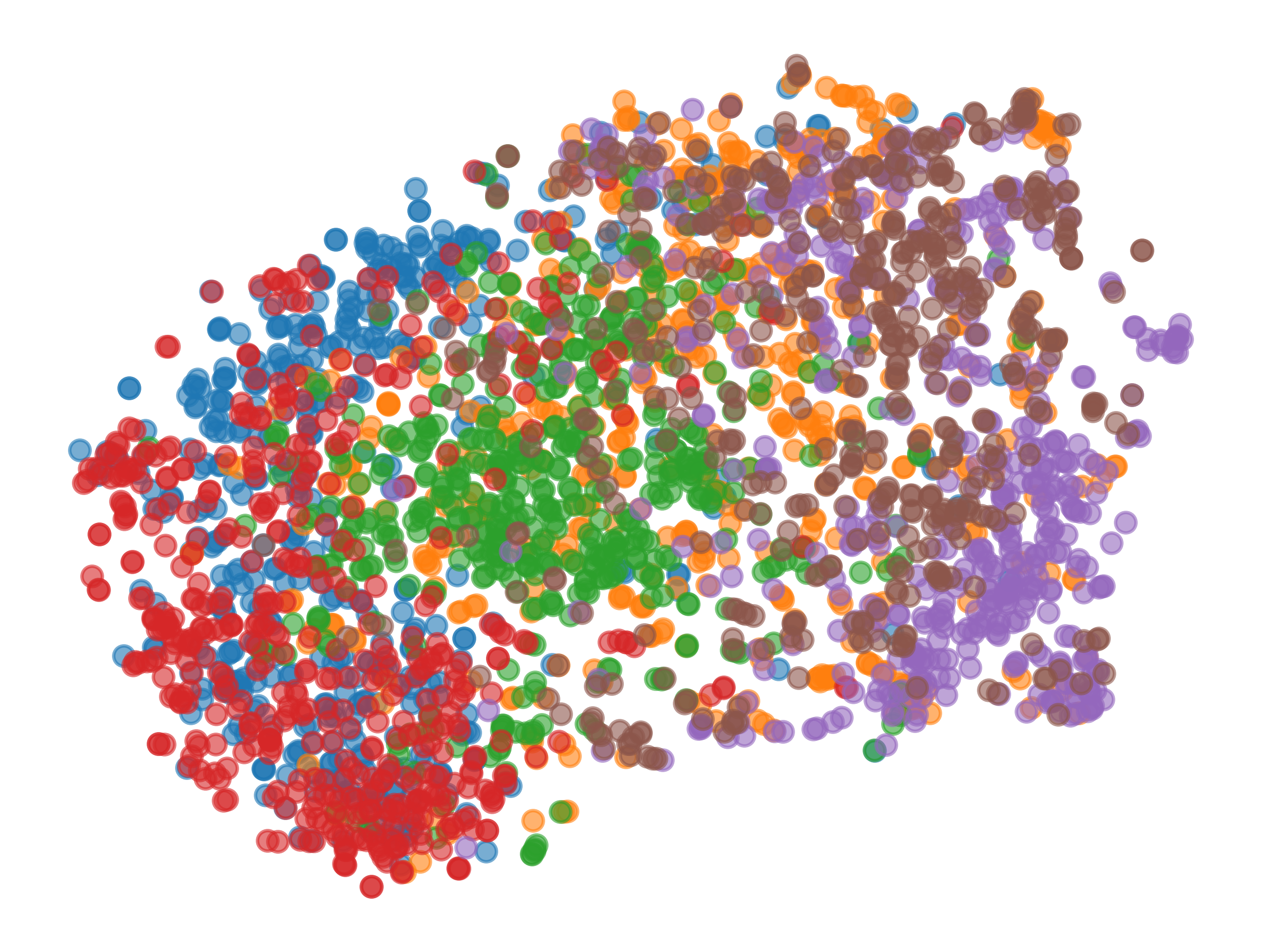}
        \caption{}
    \end{subfigure}
    \begin{subfigure}[b]{0.4\columnwidth}
        \centering
        \includegraphics[width=\textwidth]{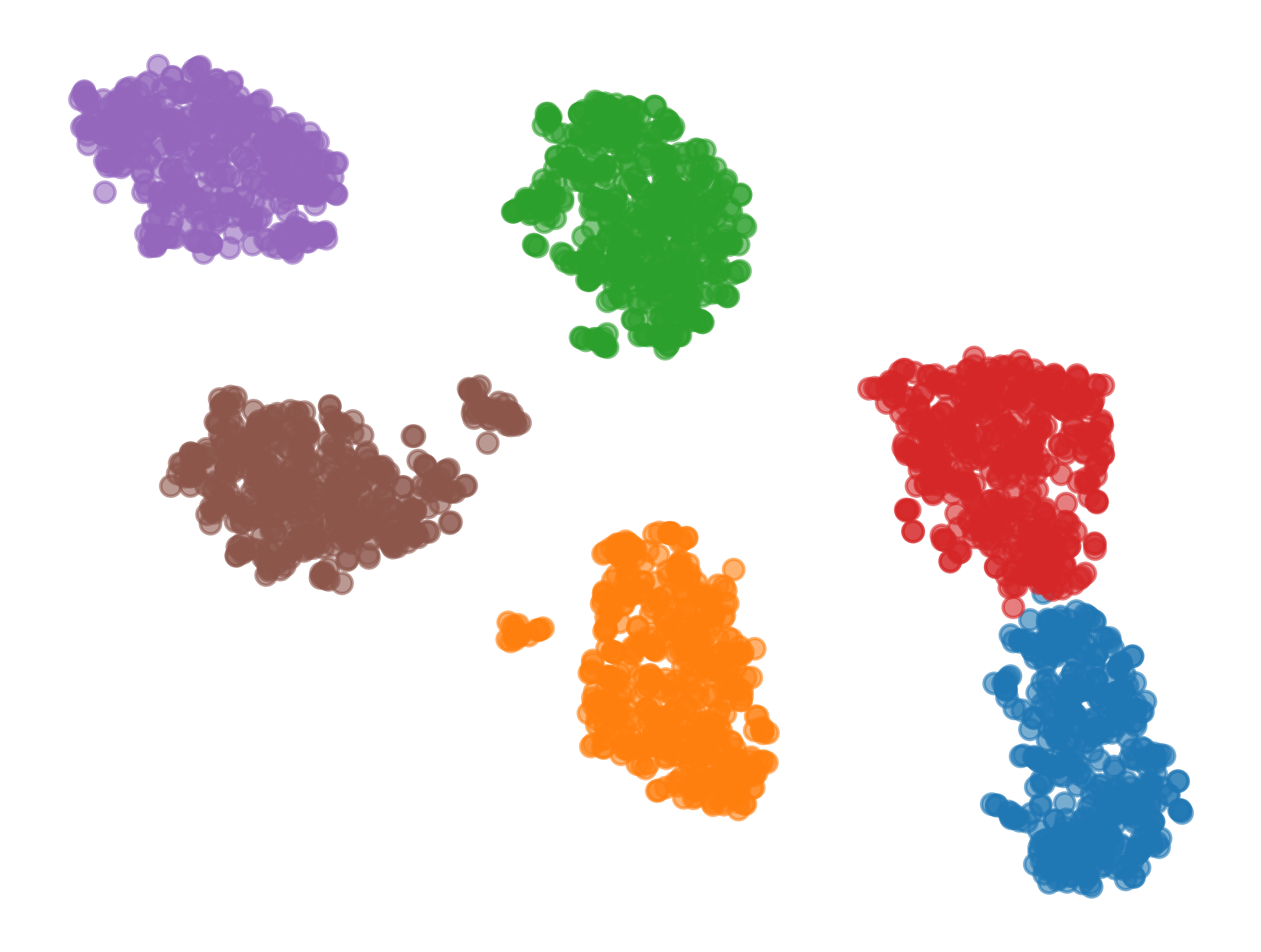}
        \caption{}
    \end{subfigure}
    \begin{subfigure}[b]{0.4\columnwidth}
        \centering
        \includegraphics[width=1.2\textwidth]{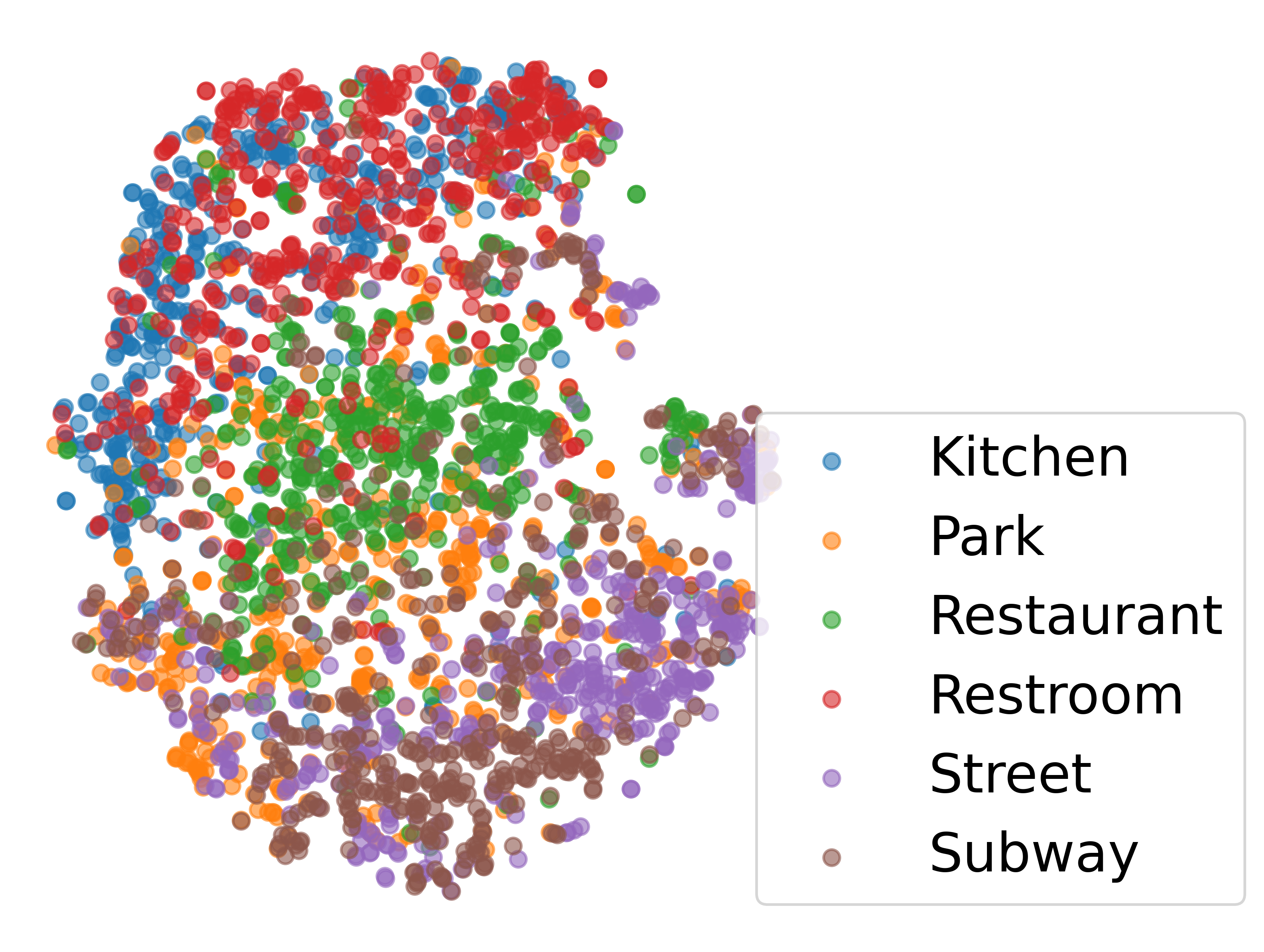} 
        \caption{}
    \end{subfigure}
    \caption{t-SNE representation of the bottleneck features from $D$ of (a) UNet, (b) $\text{UNet}_{\text{oracle}}$, and (c) $\text{UNet}_{\text{const}}$ (\emph{Embedding size II})}
    \label{fig:filmunet-oracle}
\end{figure}

As can be seen from Table~\ref{tab:results}, unconditioned UNet model obtained a decent performance, gaining 5.89 dB in SI-SDR compared to the noisy input. To explain this observation, we speculate that with the dataset where OC components are synthetically added to clean audio, the model may exploit the acoustic statistical mismatch between the two sources, and preliminary experiments have verified this possibility. Given that this is observed in UNet, we acknowledge that $\text{UNet}_{\text{ASC}}$ and $\text{UNet}_{\text{oracle}}$ could likewise have the tendency to capture the acoustic mismatch rather than primarily relying on the context for denoising. However, the improvement in the performance of the denoising models when being explicitly introduced to informative context demonstrates that conditioning on informative context enhances the denoising process. Addressing this problem of acoustic statistical mismatch would require new data collection method and is a topic for further study.

The effect of context conditioning on the denoising model's latent representations is observed in the t-SNE~\cite{van:2008:jmlr} visualizations of the bottleneck features from the denoiser $D$ in all experiments. Figures~\ref{fig:filmunet-asc} and~\ref{fig:filmunet-oracle} illustrate that, at comparable hierarchical levels, models conditioned with informative context (ASC context and Oracle context) achieve distinct clustering following the scene classes, showing that context information is retained in the model's latent representations. 
\section{Conclusion}
\label{sec:conslusion}
In this paper, we introduce the concept of \textbf{automatic audio contextual denoising} where target (what to be kept) and noise (what to be removed) depends on the learned context. For this first work, we define the context as information associated with an acoustic scene class. We propose a deep learning based method that automatically infers the context and suppresses out-of-context components relative to the learned context. Experiments are conducted on paired clean and noisy scene audio across different contexts, with out-of-context components as the noise to be removed. The proposed method is compared against threes variants: without any context inference, with oracle context, and with separately provided uninformative context. Results show that learned context consistently achieves better result than the other variants, suggesting that the model can learn the context and context-dependent processing improve denoising performance. At the same time, we acknowledge that the current dataset, with synthetically added out-of-context components to real-life recorded clean audio signal, could guide the model to learn statistical mismatch cues instead of focusing on the context to do denoising. To mitigate this confound and isolate true contextual gains, further study and experiment are needed to investigate and address this issue.

\bibliographystyle{unsrt}  
\bibliography{refs}

\end{document}